\documentclass[reqno]{amsart}
\usepackage{graphicx}% Include figure files
\usepackage{amssymb,amsmath,amsthm}
\newcommand{\bee}{\begin {eqnarray*}}
\newcommand{\eee}{\end {eqnarray*}}
\newcommand{\be}{\begin {eqnarray}}
\newcommand{\ee}{\end {eqnarray}}

\newcommand{\D}{\mathrm{d}}
\newcommand{\Real}{\mathbb{R}}
\newcommand{\C}{\mathbb{C}}
\newcommand{\pd}{\partial}

\newcommand{\Hilbert}{\mathcal{H}}

\newcommand{\eps}{\varepsilon}
\newcommand{\K}{\mathcal{K}}
\newtheorem{Lemma}{Lemma}
\newtheorem{Theorem}{Theorem}
\newtheorem{Proposition}{Proposition}

\theoremstyle{definition}
\newtheorem{Remark}{Remark}
\newtheorem{ass}{Assumption}

\begin{document}

\title []{Resonances in twisted quantum waveguides}

\author {Hynek Kova\v{r}\'{\i}k}

\address {Institute of Analysis, Dynamics and Modeling, Universit\"at
  Stuttgart, PF 80 11 40, D-70569 Stuttgart, Germany.}

\email {kovarik@mathematik.uni-stuttgart.de}

\author {Andrea Sacchetti}

\address {Dipartimento di Matematica Pura ed Applicata\\ Universit\'a
degli studi di Modena e Reggio Emilia\\ Via Campi 213/B, Modena 41100,
Italy}

\email {Sacchetti@unimore.it}

\date {\today}

\thanks{AS was partially supported by the INdAM project {\it
    Mathematical modeling and numerical analysis of quantum systems
    with applications to nanosciences} and  by MIUR under the project
  COFIN2005 {\it Sistemi dinamici classici, quantistici e
    stocastici}. \ HK is grateful to the Department of Mathematics of
  the University of Modena for the warm hospitality extended to him.}

\begin{abstract}
In this paper we consider embedded eigenvalues of a Schr\"odinger
Hamiltonian in a waveguide induced by a symmetric perturbation. \ It
is shown that these eigenvalues become unstable and turn into
resonances after twisting of the waveguide. \ The perturbative
expansion of the resonance width is calculated for weakly twisted
waveguides and the influence of the twist on resonances in a concrete
model is discussed in detail. 
\end{abstract}


\maketitle

\section {Introduction}
Quantum and electromagnetic waveguides have been studied since many
decades, see \cite {Igarashi,Lew1,Lew2,Yabe} and also \cite{LCM} where
the 
classical and the quantum pictures are compared. \ In this framework
the spectral analysis of differential operators in tubular domains 
has become a research field of a certain interest \cite {Clark,
  ES,GJ}. \ Moreover, with the introduction of nano-devices, such as
nanotubes, then new open problems in quantum transmission for such
structures appeared \cite {Cohen}.  

\begin{figure}
\begin{center}
\includegraphics[height=7cm,width=9cm]{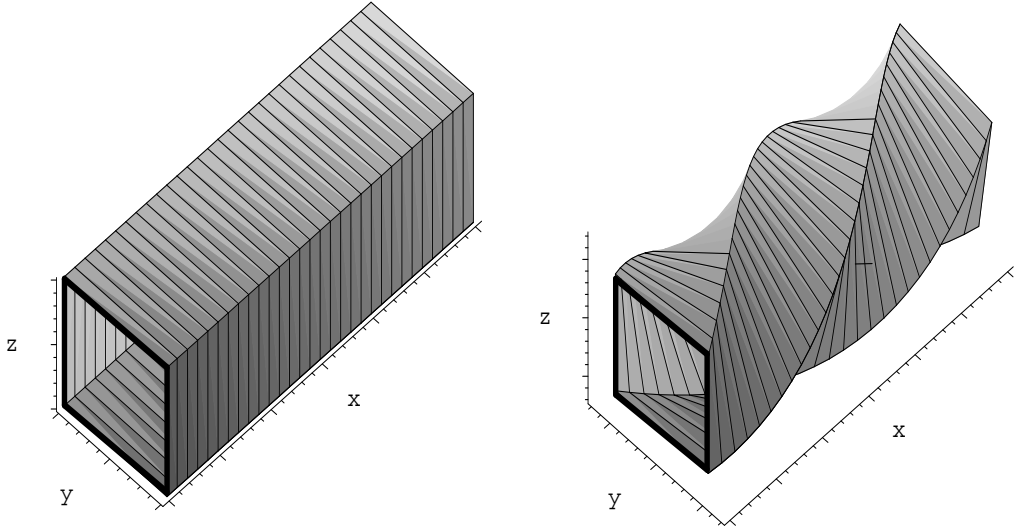}
\caption{On the left, the plot of the surface of a rectangular
  waveguide without twisting; in the right, plot of the surface of the
  twisted rectangular waveguide. \ Bold line represents the boundary
  of $\omega$.} 
\label {Fig0}
\end{center}
\end{figure}

We consider here a waveguide type domain $\Omega=\Real\times\omega$
(see Figure \ref {Fig0}, on the left), where the cross section $\omega
$ of the waveguide is an open bounded and connected subset of
$\Real^2$. \ We impose Dirichlet boundary conditions at the boundary
of $\Omega$. \ The spectrum of the free operator $-\Delta$ in
$L^2(\Omega)$ is absolutely continuous and covers the half-line
$[E_1,\infty)$, where $E_1$ is the lowest eigenvalue of the Dirichlet
Laplacian on $\omega$. \ It is a well known fact that this spectrum is
unstable against perturbations; indeed, a negative perturbation,
vanishing at infinity, of $-\Delta$ will induce at least one bound
state below the threshold $E_1$ and new embedded bound states in the
half-line $[E_1, + \infty )$ (see, e.g., Figure \ref {Fig3}). \ The
perturbation can be either of a potential type or of a geometric type,
see \cite{ChDFK,DE,GJ} and references there. Similar effects occur
also in two-dimensional strips, \cite{BEGK,BGRS,ES}. \ These new bound
states below the threshold $E_1$ correspond to the particles
(electrons) which do not propagate along $\Omega$, but remain
localized in a bounded region of $\Omega$.  

Recently it has been shown\cite{EKK} that the presence of bound states
in $\Omega$ can be, up to certain extent, suppressed by another
geometrical perturbation: the so called {\it twisting} which is
defined as follows. \ For a given $x\in\Real$ and $s:=(y,z)\in\omega$
we define the mapping  
\bee
 f_\epsilon : \Real\times\omega \to \Real^3 
\eee
by
\be 
\label{mappingL} 
f_{\eps}(x,s) = (x,\, y \cos(\eps\, \alpha (x))+z\sin(\eps\,
\alpha(x)) ,\, z\cos(\eps\, \alpha(x))-y\sin(\eps\, \alpha(x))) ,   
\ee 
where $\eps>0$ is a real parameter and $\alpha:\Real\to\Real$ is a
differentiable function. \ Furthermore, we introduce  
\bee 
\Omega_{\eps}:= f_{\eps}(\Omega) .  
\eee 
Clearly, $\Omega_{\eps}$ is a tube which is twisted unless the
function $\alpha$ is constant (in Figure \ref{Fig0} we plot,
respectively, a rectangular tube without and with twisting). \ The
result of \cite{EKK} shows that if the cross section $\omega$ is not
rotationally symmetric and the tube $\Omega$ is 
twisted, even only locally, then the bound states for the perturbed
Hamiltonian $-\Delta +V$ do not appear for any negative potential
$V(x)$, but only if $V$ is strong enough. \ In other words, one could
say that a twisting of a tubular domain $\Omega$ improves the
transport of charged particles in $\Omega$ in the sense that it
protects the particles to get trapped by weak perturbations. \ The
repulsive effect of twisting has been recently observed also
in~\cite{Gr}, where the absence of discrete eigenvalues in tubes which
are simultaneously mildly curved and mildly twisted. \ Moreover,
in~\cite{BMT} the 
repulsive effect of twisting is demonstrated for bounded tubes whose
thickness goes to zero.  

Similar results were also obtained for two-dimensional waveguides with
combined boundary conditions or with a local magnetic field
\cite{EK,KK}. \ However, the geometrical perturbations of the
waveguide generically induce also the existence of {\it resonances},
i.e. metastable states with very long lifetimes, see
\cite{DEM,DEM2,DES}.  

\vspace{0.15cm}

It is the aim of the present paper to describe the influence of
twisting on the resonances in the waveguides. \ More precisely, we
study the situation in which the free Laplacian is perturbed by an
attractive potential $V(x)$, which decays at infinity along the
waveguide direction, where $x\in \Real$ represents the coordinate
along the waveguide direction. \ The point spectrum of the perturbed
Hamiltonian $-\Delta +V(x)$ consists, in addition to the bound states
below $E_1$, of infinitely many eigenvalues embedded in the continuum
$[E_1,\infty)$ (see Figure \ref{Fig3}). \ It was shown in \cite{DEM},  

for two-dimensional waveguides, that these embedded eigenvalues
generically turn into resonances in the presence of a constant
magnetic field. \ Following the method of \cite {DEM} we show that
this happens also when the magnetic field is replaced by the twisting,
provided the cross section $\omega$ is 
not rotationally symmetric, see Theorem 1. \ For weak twisting we also
give the perturbative expansion of the corresponding resonance width.

In order to obtain a precise estimate on the imaginary part of the
resonances and, in particular, to prove that it is strictly negative
we consider   
in Section \ref{concrete} a concrete model in which the potential $V$
approximates a one-dimensional point interaction. \ For such a model
we  explicitly calculate the leading term of the imaginary part of a
chosen resonance, see Proposition \ref{example}, and we prove that for
suitable values of the parameters (see Remark \ref {notaImm}), the
imaginary part of the resonance is strictly negative.

\section{Preliminaries}\label{prelim} 

Throughout the paper we will denote by $\langle \cdot , \cdot
\rangle_H$ the scalar product in a Hilbert space $H$ with the
convention $\langle \alpha u , v
\rangle_H =\bar{\alpha} \langle u , v\rangle_H$ for all $\alpha \in\C$ and $u,v\in H$. \ 
For a
real-valued measurable bounded function $V(x)$ on $\Real$ we formally
define the Hamiltonians  
\bee
\tilde{H}_\eps^0 = - \Delta \ \ \mbox { and } \ \ \tilde{H}_\eps^V =
-\Delta +V(x) \ \ 
\mbox{in} \ \  L^2(\Omega_{\eps}) 
\eee
with Dirichlet boundary conditions at $\pd\Omega_{\eps}$. \ The
operator $\tilde{H}_\eps^V$ is associated with the closed quadratic 
form  
\be
\label{qadfrom}
\tilde{Q}^V_{\eps}[\psi] := \int_{\Omega_{\eps}}\, \left [ 
|\nabla\psi|^2\, +
  V(x) \, |\psi|^2  \right ] \D x\,\D s\, ,  
\ee
with the form domain $D(\tilde{Q}^V_{\eps })=  
\Hilbert^1_0(\Omega_{\eps})$. 

Given a test function $\psi\in C_0^{\infty}(\Omega)$ it is useful to
introduce the following shorthand,  
\be
\label{twisting}
\pd_{\tau}\psi := y\pd_{z}\psi-z\pd_{y}\psi .
\ee
As usual in such situations, in order to analyze the operator
$\tilde{H}_\eps^V$ we pass from the twisted tube $\Omega_\eps$ to the
untwisted tube $\Omega$ by means of a simple substitution of
variables. \ This gives  
\bee 
Q_{\eps}^V [\psi]=\int_{\Omega}\, \left( |\nabla_s
  \psi|^2+|\pd_x\psi+\eps\dot\alpha(x)\pd_{\tau}\psi|^2\, + V(x)\,
  |\psi|^2 \right) \D x\,\D s \, ,  
\eee 
with the form domain $D(Q^V_{\eps })=  \Hilbert^1_0(\Omega)$
and with the notation   
\bee
\nabla_s\psi:=(\pd_{y}\psi,\pd_{z}\psi)\, .
\eee
In other words, the operator $H^V_{\eps}$, associated with $Q_\eps^V$
and unitarily equivalent to $\tilde{H}_\eps^V$,
acts on its domain in $L^2
(\Omega )$  in the weak sense as  
\bee
H_{\eps}^V= -\pd_{y}^2-\pd_{z}^2-[\pd_x+\eps \dot\alpha(x)\,
\pd_{\tau}]^2+ V(x) =H_0^V + U_\eps^V  \, , 
\eee
where
\bee
H_0^V  = - \pd_{x}^2 - \pd_{y}^2-\pd_{z}^2 +V(x)
\eee
and
\bee
U_\eps^V &=& -[\pd_x+\eps \dot\alpha(x)\, \pd_{\tau}]^2 + \pd_x^2 \\ 
&=& - 2\eps\,  \dot\alpha( x)\pd_x\, \pd_{\tau} - \eps\, \ddot\alpha (
x)\, \pd_{\tau} -\eps^2 \dot\alpha^2( x)\, \pd_{\tau}^2 \, . 
\eee

\begin {Remark} \label {simmetrico}
The term $U_\eps^V$ is a symmetric operator on $L^2 (\Omega)$ with
Dirichlet boundary conditions at $\partial \Omega$. 
\end {Remark}

\noindent In order to show that the embedded eigenvalues of $H_0^V$
turn into the resonances once the waveguide is twisted, we employ the
method of the exterior complex scaling in combination with the regular
perturbation theory \cite {CFKS}. \ We start by locating the spectrum
of the  untwisted model. 

\section{Spectrum of $H_0^V$}

We will suppose that $V$ satisfies the following 
\begin{ass} \label{Ass-VA}
The function $V(x)$ is not identically equal to zero and
\be
\int_{\Real}\, (1+x^2)\, |V(x)|\, \D x< \infty \ \mbox { and } \
\int_{\Real}\, V(x) \, \D x \le 0\, . 
\ee
\end{ass}

\begin{figure}
\begin{center}
\includegraphics[height=1cm,width=8cm]{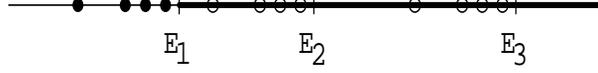}
\caption{The discrete spectrum of $H_0^V$ consists of finitely many
  simple eigenvalue below $E_1$ (denoted by full circle); the
  essential spectrum is given by the half-line $[E_1 , + \infty)$. \
  Furthermore, a non empty set of simple eigenvalues (denoted by empty
  circle) embedded in the half-line $[E_1 , + \infty)$ occurs.} 
\label{Fig3}
\end{center}
\end{figure}

\noindent It then follows from \cite {RS4} (see, e.g. Theorem XIII.110
in  and its Notes) that the operator 
\bee
h:= -\pd_x^2 +V(x) \quad \text{in}\quad L^2(\Real)
\eee
possesses finitely many negative eigenvalues $\{\mu_j\}_{j=1}^N$, $N
\ge 1$, each of multiplicity one. \ We denote by $\varphi_j (x)$ the
corresponding normalized eigenfunctions. \ The essential spectrum of
$h$ covers the positive half-line $[0,\infty)$. \ On the other hand,
it is well known that the operator $-\Delta^{\omega}_D$, i.e. the
Dirichlet Laplacian on $\omega$, is positive definite and has purely
discrete spectrum. \ Let $\{E_n\}_{n=1}^{\infty}$ be the
non-decreasing sequence of its eigenvalues and let $\chi_n (s)$ denote
the associated normalized eigenfunctions. \ The set of such 
eigenfunctions is an orthonormal basis of $L^2 (\omega )$. \ We denote 
by
\bee
\Sigma = \left \{ E= \mu_j + E_n , \ j=1,\ldots ,N , \ n \ge 1 \right 
\}
\eee
the set of eigenvalues of $H_0^V$ with associated normalized 
eigenvectors
\bee
\psi_{n,j}(x,s) = \varphi_j (x) \chi_n (s),
\eee
and
\bee
\Sigma_+ = \Sigma \cap [E_1 , +\infty ) ,\ \Sigma_- = \Sigma \cap (-
\infty, E_1 ) 
\eee
where $\Sigma_-$ is not empty since $\mu_j <0$ for any $j$. \ Then, by
the standard arguments, \cite{RS4}, the spectrum of 
\bee
H_0^V = -\Delta +V(x)\,, \quad {\rm in\, \, } L^2(\Real\times\omega)
\eee
is given by $\sigma(H_0^V) = \sigma_d(H_0^V) \cup \sigma_{ess}(H_0^V)$, 
where
\bee
\sigma_d(H_0^V) = \Sigma_- \ \mbox { and } \ \sigma_{ess}(H_0^V) =
[E_1,\infty) . 
\eee
In addition, $H_0^V$ possesses point spectrum embedded into the
continuum given by $\Sigma_+$ (see Figure \ref {Fig3}). 

\vspace{0.15cm}

\noindent
We expect that when $\eps$ becomes non-zero then these embedded
eigenvalues generically turn into resonances, which are the main
object of our study. 

\begin {Remark} \label {reale}
Since the operator $H_0^V$ commutes with complex conjugation then its
eigenfunctions $\psi$ can be assumed to be real-valued. 
\end {Remark}

\section{Complex scaling}
\label{Smain}

Henceforth, we will employ the method of exterior complex scaling to
the operator $H_\eps^V$. \ In order to do so, we will need some
assumptions on the functions $V$ and $\alpha$: 

\begin{ass} \label{Ass-VB}
$V$ extends to analytic function with respect to $x$ in some sector
\bee
M_{\beta}:= \{\zeta \in\C\, :  |\arg \zeta |\leq \beta\}, \ \mbox {
  with } \ \beta >0. 
\eee
Moreover, $V$ is uniformly bounded in $M_{\beta}$.
\end{ass}

\begin {ass} \label {Ass-VC}
$\alpha$ extends to analytic function with respect to $x$ in
\bee
{\mathcal M}_\beta = M_{\beta} \cup \left \{ \zeta \in \C \ :\ |\Im
  \zeta | \leq \beta \right \}, \ \mbox { with } \ \beta >0. 
\eee
and $\dot \alpha$ is uniformly bounded in ${\mathcal M}_{\beta}$. \ In
addition $\dot \alpha (x) > 0 , \ \forall x \in \Real  $. 
\end{ass}

\begin {Remark}
Since $\dot \alpha$ is uniformly bounded in ${\mathcal M}_{\beta}$
then from the Cauchy theorem it follows that $\ddot \alpha $ is
uniformly bounded in ${\mathcal M}_{\beta'}$ for any $0<\beta' <
\beta$. 
\end {Remark}

\noindent In analogy with \cite{DEM} we introduce the mapping
$S_\theta$, which acts as a complex dilation in the longitudinal
variable $x$: 
\bee
(S_{\theta}\psi)(x,s) = e^{\theta/2}\psi(e^{\theta} x,s)\, , \quad
\theta \in \C\, . 
\eee
The transformed operator then takes the form
\bee
H_{\eps}^V (\theta ) = S_{\theta} H_{\eps}^V S^{-1}_{\theta} = H_0^V
(\theta ) + U_\eps^V (\theta)\, ,  
\eee
where
\bee
H_0^V (\theta ) = S_{\theta} H_{0}^V S^{-1}_{\theta} = -e^{-2\theta}
\, \pd_{x}^2 - \pd_{y}^2-\pd_{z}^2 +V(e^{\theta}x) \, 
\eee
and
\be \label{perturbation}
U_\eps^V (\theta ) = S_{\theta} U_{\eps}^V S^{-1}_{\theta} = - 2\eps\,
e^{-\theta}\,  \dot\alpha( e^\theta x)\pd_x\, \pd_{\tau} - \eps\,
e^{-\theta}\, \ddot\alpha ( e^\theta x)\, \pd_{\tau} -\eps^2
\dot\alpha^2( e^\theta x)\, \pd_{\tau}^2 \, . 
\ee

\begin{figure}
\begin{center}
\includegraphics[height=2cm,width=8cm]{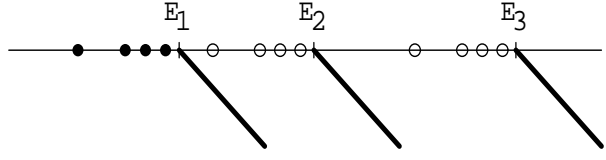}
\caption{The discrete spectrum of $H_0^V (\theta )$ consists of a
sequence of real and simple eigenvalues (denoted by circle); the
essential spectrum is given by the half-lines $E_n + e^{-2i\Im
\theta } \Real^+$.} 
\label {Fig4}
\end{center}
\end{figure}

\begin {Lemma}
Let $V$ satisfy assumptions \ref {Ass-VA} and \ref {Ass-VB}, then
$H_{0}^V (\theta )$ is an analytic family of type {\rm A} with respect
to $\theta$. \ Furthermore, the spectrum of $H_{0}^V (\theta )$ has
the form (see Figure \ref {Fig4}) 
\be \label{sum}
\sigma \left ( H_{0}^V (\theta ) \right ) =  \bigcup_{n} \left [ E_n +
  e^{-2i\Im \theta } \Real^+ \right ]. 
\ee
More precisely, the essential spectrum of $H_0^V (\theta )$ consists
of the sequence of the half-lines $E_n + e^{-2i\Im \theta } \Real^+ $,
$n=1,2,\ldots $, and the discrete spectrum of $H_0^V (\theta )$
consists of the set of eigenvalues $\mu_j + E_n$ with associated
eigenvectors 
\be
\left [ \psi_{n,j} (\theta ) \right ] (x,s) = \left [ S_\theta
  \psi_{n,j} \right ] (x,s) = e^{\theta /2} \varphi_j (e^\theta x )
\chi_n (s). \label {vet1} 
\ee
\end {Lemma}

\begin {proof}
It follows from Assumption \ref{Ass-VB} that the family of operators
$H_0^V(\theta)$ in analytic of type A with respect to $\theta$, see 
\cite[Chap.7]{Kato}. \ For what concerns its spectrum it is enough to
remark that the operator 
\bee
h (\theta ) = S_\theta h S_\theta^{-1} = -e^{-2\theta} \, \pd_{x}^2
+V(e^{\theta}x) 
\eee
in $L^2 (\Real )$ has the spectrum given by
\bee
\sigma (h(\theta )) = \left \{ \mu_1 , \ \ldots , \ \mu_N \right \}
\cup e^{-2i\Im \theta } \Real^+ \, . 
\eee
By the Ichinose's Lemma, see \cite[Sec.~XIII.10]{RS4}, we then 
obtain
the formula \eqref{sum} for the spectrum of the sum $h(\theta)
-\pd_y^2-\pd_z^2$.  
\end {proof}

\begin {Lemma} \label{analytic}
Let $V$ satisfy assumptions \ref {Ass-VA} and \ref {Ass-VB} and let
$\alpha$ satisfy assumption \ref {Ass-VC}, then the operator $U_\eps^V
(\theta)$ is a relatively bounded perturbation of $H_{0}^V  (\theta
)$. \ Moreover, the family of operators $H_\eps^V (\theta )$ is
analytic of type A for all $\theta$ such that $|\theta| < R_\eps$,
where $R_\eps\to\infty$ as $\eps\to 0$. 
\end {Lemma}

\begin {proof} 
In order to prove the Lemma we consider a test function $\psi\in
C_0^{\infty}(\Omega)$. \ Using the assumptions on $V$ we find out that
there exists a positive constant $c$ (here and below, $c$ will denote
a positive constant whose value changes from line to line) such that  
\bee
\|H_0^V(\theta)\, \psi\|^2 \geq c\,
(\|\pd_x^2\psi\|^2+\|\pd_y^2\psi\|^2+\|\pd_z^2\psi\|^2)-c \|\psi\|^2\,
.  
\eee
From this inequality and using the fact that $\omega $ is bounded we 
arrive at
\bee
\|\pd_{\tau}^2\psi\|^2 &\leq & c\,
(\|\pd_z^2\psi\|^2+\|\pd_y^2\psi\|^2+\|\pd_z\psi\|^2 +\|\pd_y\psi\|^2)
\\   
&\le & c  (\|\pd_z^2\psi\|^2+\|\pd_y^2\psi\|^2+\|\psi\|^2 ) \\ 
&\le & c  (\| H_0^V (\theta ) \psi \|^2+\|\psi\|^2 )
\eee
and
\be \label{relbound1}
\| \pd_\tau \psi\|^2 \leq c (\| \pd_z \psi \|^2 +\| \pd_y \psi \|^2)
\leq c\, \left ( \| H_0^V (\theta) \, \psi \|^2 +\| \psi \|^2 \right
)\, .  
\ee 
As for the mixed term in $U_\eps^V (\theta )$ we note that
\be
\|\pd_x \pd_\tau \psi\|^2 = \left \langle \pd_x\pd_{\tau}\psi,
  \pd_x\pd_{\tau}\psi \right \rangle_{L^2(\Omega )} \leq
\|\pd_{\tau}^2\psi\|\, \|\pd_x^2\psi\| \leq \frac 12 \left [
  \|\pd_{\tau}^2\psi\|^2+\|\pd_x^2\psi\|^2 \right ] \, . 
\ee
Collecting all these estimates we finally conclude that there exists a
positive constant $C$, depending on $\omega$ and $V$, such that 
\be \label{relbound2}
\|U_\eps^V (\theta ) \, \psi\|^2 \leq C\, (\eps+\eps^2 )^2\, (
\|H_0^V(\theta)\, \psi\|^2+\|\psi\|^2)\, . 
\ee
To prove the second statement of the Lemma we first notice that by
assumption \ref{Ass-VB} we have $D(H_0^V ({\theta}))=D(H_0^V (0))$. \
By assumption \ref{Ass-VC} and \cite[Sec. 7.2]{Kato} it thus suffices
to show that both $\pd_x\pd_{\tau}$ and $\pd_{\tau}$ are relatively
bounded with respect to $H_0^V ({\theta})$. \ However, this follows
from \eqref{relbound1} and  \eqref{relbound2}. \end {proof} 

\noindent
Lemma \ref{analytic} tells us that the eigenvalues of $H_\eps^V
(\theta )$ are analytic functions of $\theta$. \ By a standard
argument, \cite{CFKS}, it turns out, that they are in fact independent
of $\theta$. \ The non-real eigenvalues of $H_\eps^V (\theta )$, for
$\theta$ such that $\Im \theta >0$, are identified with the resonances
of $H_\eps^V$ \cite{CFKS}.  

\begin {Remark}
As a result of the previous proof it follows that $U_\eps^V (\theta )$
is a regular perturbation of the operator $ H_0^V ({\theta})$. \ This
enables us to apply the analytic perturbation theory to the
eigenvalues of the operator $ H_0^V ({\theta})$. 
\end {Remark} 

\begin {Theorem} \label{main}
Let $E =E_n + \mu_j \in \Sigma_+$ be a simple eigenvalue of $H^V_0
\theta )$ where $\theta$ is a fixed complex number such that $\Im
\theta >0$. \ For any ball $B$ centered in $E$ there exists
$\eps^\star >0$, depending on $E$, such that for any $\eps$ with
$|\eps| < \eps^\star$, there is an eigenvalue $E(\eps )$ of $H^V_\eps
(\theta )$ belonging to $B$ and with the imaginary part given by  
\be
\Im E(\eps ) = - \eps^2 a + O(\eps^3 ) \label {imma}
\ee
where $a $ is a constant independent of $\eps$ and equal to
\be \label{a}
a= \sum_{k \leq k^\star} \left | \langle \pd_\tau \chi_n , \chi_k
  \rangle_{L^2 (\omega )} \right |^2 \left \langle v_j , \Im \hat r
  (E-E_k) v_j \right \rangle_{L^2 (\Real)} \label {immabis}\, . 
\ee
Here
\bee
v_j  = (-2 \dot \alpha \pd_x + \ddot \alpha ) \varphi_j , \ k^\star
=\max \left \{ k \ : \ E_k - E < 0 \right \} 
\eee
and $\Im \hat r$ stands for the imaginary part of the reduced
resolvent \cite {Kato} of $h = - \pd_x^2 + V$ with respect to the
eigenvalue $\mu_j$: 
\bee
&& \left \langle v_j , \Im \hat r (E-E_k) v_j \right \rangle_{L^2 
(\Real)} = \\
&& \ \  \lim_{\rho \to 0^+} \frac {1}{2i} \left [ \left \langle v_j ,
    \hat r (E-(E_k+i\rho )) v_j \right \rangle_{L^2 (\Real)} - \left
    \langle v_j , \hat r (E-(E_k-i \rho )) v_j \right \rangle_{L^2
    (\Real)} \right ] , 
\eee
\end {Theorem}

\begin {proof}
Let $\psi (\theta )= \psi_{n,j}(\theta )$ be the associated normalized
eigenvector (\ref {vet1}) belonging to $E$, where $n$ and $j$ are
fixed. \ We apply the regular perturbation theory saying that for some
fixed $r>0$ small enough and for any $\eps$ with modulus small enough,
in the given ball $B_r (E)$ exists only one eigenvalue $E(\eps )$ of
$H_\eps^V (\theta )$ with associated eigenvector  
\bee
\psi^\eps (\theta ) = \frac {1}{2\pi i} \oint_{\partial B_r} \left
  [\zeta - H_\eps^V (\theta )\right ]^{-1} \psi (\theta ) \D \zeta \,
. 
\eee
Furthermore, the regular perturbation theory also yields that this
eigenvalue is given by means of the convergent perturbative series
(see, e.g. \cite[XII.6]{RS4})  
\bee
E(\eps )= \frac {\left \langle \bar \psi (\theta ) , H_\eps^V (\theta
    ) \psi^\eps (\theta ) \right \rangle_{L^2 (\Omega )} }{\left
    \langle \bar \psi (\theta ) ,  \psi^\eps (\theta ) \right
  \rangle_{L^2 (\Omega )} } = \sum_{m=0}^\infty e_m (\eps ) , \  e_m =
O(\eps^m ) 
\eee
where, as usual,
\bee
e_0 =E \ \mbox { and } \ e_1 = \frac {\langle \bar \psi (\theta ) ,
  U_\eps^V (\theta ) \psi (\theta ) \rangle_{L^2 (\Omega )} }{\langle
  \bar \psi (\theta ) , \psi (\theta ) \rangle_{L^2 (\Omega )} } =
\frac {\langle \psi , U_\eps^V \psi \rangle_{L^2 (\Omega )} }{\langle
  \psi ,  \psi \rangle_{L^2 (\Omega )} } 
\eee
are constant with respect to $\theta$; $\psi$ is the real-valued
vector \eqref {vet1} for $\theta=0$ (see Remark \ref {reale}). \ These
constants $e_0$ and $e_1$ are real-valued since $U_\eps^V$ is a
symmetric operator (see Remark \ref {simmetrico}). \ If we prove that
$\Im e_2 = - \eps^2 a + O(\eps^3 )$ for some $a>0$ independent of
$\eps$ then the stated result follows. \ To this end we recall that
(see \cite {DEM})  
\bee
\Im e_2 = \Im a_2 \left (1+ O(\eps ) \right )\, ,
\eee
where
\bee
a_2 &=& - \frac {1}{2\pi i}\oint_{\partial B_r} \left \langle \bar
  \psi (\theta ) , U_\eps^V (\theta ) \left [\zeta - H_0^V (\theta
    )\right ]^{-1} U_\eps^V (\theta ) \psi (\theta ) \right
\rangle_{L^2 (\Omega )} \frac {\D \zeta}{\zeta - E} \\ 
&=& \lim_{\rho \to 0^+} f(\theta , E+i\rho ) = \lim_{\rho \to 0^+}
f(\theta =0 , E+i\rho ) 
\eee
and
\bee
f(\theta , \zeta ) &=& - \left \langle \bar \psi (\theta ) , U_\eps^V 
(\theta ) \left [ \zeta - H_\eps^V (\theta ) \right ]^{-1} U_\eps^V 
(\theta ) \psi (\theta ) \right \rangle_{L^2 (\Omega )} + \\ 
&& \ \ + \left | \left \langle \bar \psi (\theta ) , U_\eps^V (\theta ) 
\psi (\theta ) \right \rangle_{L^2 (\Omega )} \right |^2 (\zeta - E 
)^{-1}.
\eee
Hence
\be
a_2 = - \langle \psi , U_\eps^V \hat R (E+i0) U_\eps^V \psi 
\rangle_{L^2 (\Omega )} := \lim_{\rho \to 0^+} \left [ - \langle \psi , U_\eps^V 
\hat R (E+i\rho ) U_\eps^V \psi \rangle_{L^2 (\Omega )} \right ] 
\label {wip}
\ee
where $\hat R(\zeta ) $ is the reduced resolvent of $H_0^V$ with
respect to the eigenvalue $E$, see \cite[X.II.6]{RS4} and \cite
{Kato}. \ Recalling that $\psi$ has the form  
\bee
\psi (x,s)= \psi_{n,j} (x,s)= \varphi_j (x) \chi_n (s)
\eee
for some $n$ and $j$ and that $\left \{ \chi_k (s) \right
\}_{k=1}^\infty$ is a basis of $L^2 (\omega )$, we obtain  
\bee
U_\eps^V \psi(x,s) = \sum_{k=1}^\infty d_k (x)  \chi_k(s), \ \mbox {
  where } \ d_k (x) = \langle \chi_k , U_\eps^V \psi \rangle_{L^2
  (\omega)}\, . 
\eee
Using the fact that $U_\eps^V \psi\in L^2(\Omega)$, Lemma
\ref{analytic}, and the dominated convergence theorem we conclude that  

\be 
\label{atwo}
a_2 = - \sum_{k=1}^\infty \left \langle d_k  , \hat r (E-E_k +i0) d_k
\right \rangle_{L^2 (\Real )}  \, ,   
\ee
where $\hat r (\zeta )$ is the reduced resolvent of $h = -
\partial_x^2 + V$ with respect to $\mu_j$. \ Concerning the imaginary
part of $a_2$ we point out that only finitely many terms on the
r.h.s.~of \eqref{atwo} have a non zero imaginary part. \ The latter
follows from the fact that $\langle d_k , \hat r (E-E_k +i0) d_k
\rangle$ is real for any $k$ large enough; more precisely for any $k >
k^\star$, where  
\bee
k^\star =\max \left \{k\, :\,  E_k - E < 0 \right\} \, .
\eee
From
\bee
d_k (x)  = \varepsilon v_j(x) \langle \pd_\tau \chi_n , \chi_k
\rangle_{L^2 (\omega )} \left [1+O(\eps )\right ], \quad v_j  = (-2
\dot \alpha \pd_x + \ddot \alpha ) \varphi_j 
\eee
we can thus conclude that
\bee
a_2 = - \eps^2 A \left [ 1 + O(\eps ) \right ]\, , 
\eee
where
\bee
A = A_{n,j} =\sum_{k \leq k^\star} \left | \langle \pd_\tau \chi_n ,
  \chi_k \rangle_{L^2 (\omega )} \right |^2 \left \langle v_j , \hat r
  (E-E_k +i0) v_j \right \rangle_{L^2 (\Real )}  
\eee
is independent of $\eps$. \ This implies (\ref {immabis}) since
\bee
a = \Im A = \sum_{k \leq k^\star} \left |\langle \pd_\tau \chi_n ,
  \chi_k \rangle_{L^2 (\omega )} \right |^2 \left \langle v_j , \Im
  \hat r (E-E_k) v_j \right \rangle_{L^2 (\Real )} . 
\eee
\end {proof}

\begin {Remark} 
Notice that if $\omega$ is rotationally symmetric, then $a=0$. \
Indeed, since $\chi_n$ is rotationally symmetric whenever $E_n$ is
simple, this  
follows from \eqref{a}. 
\end {Remark}

\begin{figure}
\begin{center}
\includegraphics[height=2cm,width=8cm]{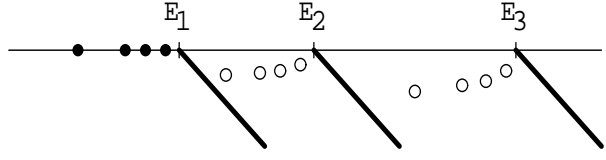}
\caption{The discrete spectrum of $H_\eps^V (\theta )$, for $\eps
  \not= 0$ small enough, consists of two parts; the first part is
  given by the real and simple eigenvalues (full circle) below $E_1$,
  the second one is given by simple eigenvalues with real part larger
  that $E_1$ and with \emph {imaginary part strictly negative} (empty
  circle).}  
\label {Fig5}
\end{center}
\end{figure}

\begin {Remark}
We point out that $\Im r (\zeta )$ is a symmetric and positive
operator for $\zeta$ real (see, e.g., \cite {DEM}). \ We can thus
\emph {generically} expect that for any ${\tilde E}>E_1$ fixed there
exists $\eps^\star >0$ small enough such that $H_\eps^V (\theta )$
does not have discrete spectrum in the interval $[E_1 , {\tilde E}]$
for any $0<|\eps | \leq \eps^\star$; more precisely, for any $\delta
>0$ the set   
\bee
\sigma_d (H_\eps^V (\theta )) \cap \left \{ [E_1 , {\tilde E}] \times
  i [-\delta ,+\delta ] \right \}   
\eee
is empty or it consists of finitely many points with imaginary part
strictly negative (see Fig. \ref {Fig5}). \ As a result it follows
that the embedded eigenvalue $E\in \Sigma_+$ of the untwisted model
turns into a resonance when an appropriate twisting is applied.  

Concerning the assumption on the multiplicity of $E$ we note
  that it is closely related to the multiplicity of $E_n$, the 
eigenvalues of
  $-\Delta^\omega_D$. In general, and
  especially for cross-sections
  with a high degree of symmetry, some eigenvalues of
  $-\Delta^\omega_D$ might be higly degenerated. However, 
this degeneracy is unstable under small perturbations of the domain,
\cite{U}. \ In fact, if the
  boundary of $\omega$ can be $C^k-$smoothly embedded in $\Real^2$ 
with $k>4$, then the eigenvalues $E_n$ of $-\Delta^\omega_D$ are {\it
  generically} simple, see \cite{U}.   
\end {Remark}

\begin {Remark}
In \cite[Thm.~1]{Gr} Grushin has obtained a similar asymptotic
expansion for the ground state in mildly twisted tubes with
transversal potential. \ He 
then proved, under certain assumptions on $\omega$, the absence of
discrete eigenvalues for $\eps$ small enough. \ We would like to
mention that there is one important difference between our result and
that of \cite{EKK}, \cite{Gr}. \ Assume that the boundary of $\omega$
is sufficiently regular (e.g.~$C^1-$smooth) and replace the Dirichlet
boundary condition at $\pd\Omega$ by the {\it Neumann} one, let us
denote the resulting operator by $H_\epsilon^{V,N}$ \ Since $V$
depends only on $x$, the eigenfunction associated with the lowest
eigenvalue of $H_0^{V,N}$ will be given by $\psi_1(x,s) = c\,
\varphi_1(x)$, where $c$ is a constant. \ This follows from the fact
that $\chi_1$ is constant in the case of Neumann boundary
conditions. \ Consequently we have $\pd_\tau\, \psi_1\equiv 0$ and
therefore, using $\psi_1$ as a test function, we find out that    
\bee
\inf \sigma(H_\eps^{V,N}) \leq \inf\sigma(H_0^{V,N})\, .
\eee
This means that, contrary to the Dirichlet case, the lowest eigenvalue
of $H_0^{V,N}$ will not be removed by the twisting, even for very
small $V$. \ On the other hand, the above analysis shows that the
effect on the embedded eigenvalues will typically occur also in the
Neumann case, since the eigenfunctions of $H_0^{V,N}$ associated with
the embedded eigenvalues are not constant in $s$. 
\end {Remark}

\section {One concrete model}
\label{concrete}

In the previous section we have seen that the embedded eigenvalues
under the influence of twisting generically turn into resonances. \
However, Theorem \ref{main} does not a priori say that the imaginary
part of the resonances is strictly negative. \ In this section, we
will show on a concrete model that for mildly twisted waveguides one
can guarantee the negativity of the imaginary part of a chosen
resonance.     

To make this problem simpler we would like to consider a concrete
model, in which $V$ acts as a Dirac delta potential. \ However, as the
Dirac delta potential is obviously not dilation analytic, see
Assumption \ref{Ass-VB}, we will approximate it by the sequence  
\be \label{cosh}
V_\nu(x) = -\frac {\nu}{2 \cosh^2 ( \nu x)}, \qquad \nu >0,
\ee
which converges to the delta function at zero as $\nu\to\infty$ in the
sense of distributions. \ Moreover, to be able to give some quantitave
results we assume that $\alpha (x)=x$. 

\begin{Proposition} \label{example}
Let $\alpha (x)=x$ and assume that the embedded eigenvalue $E=
E_2+\mu_1$ of the operator $H^{V_\nu}_0$ is simple and that  $E_2-E_1
> \frac 14$. \ Here $V_\nu$ is given by \eqref{cosh}. \ Then in the
vicinity of $E$ there is an eigenvalue $E(\eps ,\nu)$ of
$H_{\eps}^{V_\nu}(\theta)$ with the imaginary part 
given by  
\be \label{additional}
\Im E(\eps ,\nu )= -\eps^2 a (\nu ) + O(\eps^3)\, ,
\ee
where
\be
\label {limite}
\lim_{\nu\to\infty}\, a(\nu ) =  |C_1|^2\,\, \frac {\sqrt {E_2-E_1
    -\frac 14 }}{ (E_2-E_1)^2 } \, , \ C_1 = \langle \chi_1 ,
\pd_\tau\, \chi_2 
\rangle_{L^2(\omega)}\, .
\ee
In particular, if $C_1 \not=0$ then the coefficient $a (\nu )$ is
strictly positive for $\nu$ large enough.  
\end{Proposition}

\begin {Remark}
We note that with this choice of $\alpha$ the essential spectrum 
of $H_\eps^V(\theta)$ will depend on $\eps$. \ This fact is not usual in complex scaling methods; however this inconvenience does not affect the existence of the resonances since the correction of the essential spectrum is small for $\epsilon$ small and thus the non-real eigenvalues of $H_\eps^V(\theta)$ will stay far enough from its essential spectrum. \ In fact, it could be possible to avoid this fact by choosing 
\be
\label {alpha}
\alpha (x) =  x\, ,\ \forall x \in [-X , X ] \ \mbox { and } \ \dot \alpha (x) =0\, , \ \forall |x| > 2 X,
\ee
for some $X\gg 1$. \ We note that with this choice of $\alpha$ the essential spectrum 
of $H_\eps^V(\theta)$ will not depend on $\eps$ and, furthermore, since the wavefunction $\varphi_1$ decays very fast as $|x|$ grows then, in order to compute (\ref {imag}) and (\ref {dkbis}), we can practically take $\dot \alpha =1$ and $\ddot \alpha =0$.
\end {Remark}

\begin{Remark} \label {notaImm}
Equation \eqref{limite} implies that if $C_1 \neq0$, then the
imaginary part $\Im E(\eps ,\nu )$ will be strictly negative for
suitable values of $\nu$ end $\eps$. Indeed, first we take $\nu$ large
enough so that $a(\nu)>0$. \ Then we keep this $\nu$ fixed and take
$\eps =\eps(\nu )$ small enough so that the remainder term
$O(\eps^3)$, which also depends on $\nu$, becomes smaller than $\eps^2
a(\nu)$. \ Then    
\be
\Im E(\eps , \nu ) <0, \label {stima}
\ee
which means that the twisting pushes the eigenvalue $E(\eps,\nu )$
down in the complex plane, making thus the lifetime of the
corresponding resonance shorter.  
\end{Remark} 

\begin{Remark}
Note that the assumption $E_2-E_1 > \frac 14$ is made only for the
sake of simplicity. \ It guarantees that $E$ is an embedded eigenvalue
of $H^{V_\nu}_0$ for any positive $\nu $. \ If $E_2-E_1 < \frac 14$,
then we would have to consider an eigenvalue $\tilde{E}= E_k+\mu_1$
for some $k$ large enough such that $\tilde{E}$ is embedded.  
\end{Remark}

\begin{Remark} 
The coefficient $C_1$ depends on the geometry of $\omega$. \ An
integration by parts shows that $C_1= -\langle \pd_\tau\, \chi_1 ,
\chi_2 \rangle_{L^2(\omega)}$. \ It is therefore easy to see that
$C_1=0$ for rotationally symmetric $\omega$. \ However, it is not a
priori guaranteed that $C_1\neq 0$ whenever $\omega$ is not
rotationally symmetric, although we are not aware of any
counter-example. \ Let us only mention that $C_1\neq 0$ for certain 
domains $\omega$. Indeed, for $\omega=[0,a]\times[0,b]$ with $a>b>0$, 
we have
\bee
\chi_1 (y,z) = \frac{2}{\sqrt {ab}}\, \sin \left (\frac{\pi}{a}\, y
\right ) \sin \left (\frac{\pi}{b}\, z \right ) \ \ , \ \ \chi_2 (y,z)
= \frac{2}{\sqrt {ab}}\, \sin \left (\frac{2\pi}{a}\, y \right ) \sin
\left (\frac{\pi}{b}\, z \right ) \, .  
\eee
An explicit calculation then shows that
\bee
C_1 = \langle \chi_1 , \pd_\tau\, \chi_2 \rangle_{L^2(\omega)} =
-\frac{4b}{3a} \neq 0\, .  
\eee
\end{Remark}

\subsection{Proof of Proposition \ref{example}}

Equation (\ref{additional}) follows directly from Theorem \ref
{main}. \ The rest of the proof will be given in two steps.  

\subsubsection* {Spectrum of $h_\nu = - \pd_x^2 + V_\nu$}

Following \cite[\S 23]{LL} we set
\be
t = \frac 12 \left [ -1 + \sqrt {1+ \frac {2}{\nu}} \right ]\, .
\ee
The eigenvalue problem $h_\nu \tilde \varphi_j = \mu_j \tilde
\varphi_j$ admits solutions 
\be
\mu_j = - \frac {\nu^2}{4} \left [ - (2j-1)+\sqrt {1+ \frac {2}{\nu}}
\right ]^2 , \ 1\le j < t+1 
\ee
with associated eigenfunctions
\be
\tilde \varphi_j (x) = (1-\xi^2)^{e_j/2} {\boldmath F} \left [ e_j -t,
  e_j +t+1, e_j+1, \frac 12 (1-\xi ) \right ], 
\ee
where
\be
\xi = \tanh (\nu x) , \ e_j =\frac {\sqrt {-\mu_j}}{\nu} = \frac 12
\left [ - (2j-1)+\sqrt {1+ \frac {2}{\nu}} \right ], 
\ee
$\boldmath F$ denotes the hypergeometric function and $e_j - t
=j-1$. \ In particular, when $\nu \gg 1$ then $t \sim \frac {1}{2\nu}
\ll 
1 $ and the spectrum of $h$ consists of only one eigenvalue
\be
\mu_1 = - \frac {\nu^2}{4} \left [ - 1+\sqrt {1+ \frac {2}{\nu}}
\right ]^2 \sim - \frac {1}{4} + O (\nu^{-1}) \label {autoval0} 
\ee
with the associated normalized eigenvector
\be
\varphi_1 (x) = \frac {\tilde \varphi_1 (x)}{\| \tilde \varphi_1 (x)
  \|_{L^2(\Real )}} , \ \tilde \varphi_1 (x) = \left [ 1-\tanh^2 (\nu
  x) \right 
]^{e_1/2} .  \label {eigenv0}
\ee
We recall that the absolute continuous spectrum of the operators 
\bee
h_\nu = - \pd_x^2 + V_\nu (x)\quad \text{and}\quad h_\infty = -
\pd_x^2 - \delta, 
\eee 
where $\delta$ denotes the Dirac's delta at $x=0$, coincides with the
positive real axis: 
\bee
\sigma_{ac} (h_\nu ) = \sigma_{ac} (h_\infty ) = [0,+\infty )
\eee
and that, see e.g. \cite[Thm.~3.2.3]{Al}, $h_\nu \to h_\infty $ as
$\nu \to + \infty$ in the norm resolvent sense: 
\bee
\lim_{\nu \to \infty } \left \| r_\nu (\zeta ) - r_\infty (\zeta )
\right \| =0 ,\ \ \Im \zeta >0 ,  
\eee
where $r_\nu (\zeta )= \left [\zeta - h_\nu \right ]^{-1}$ and
$r_\infty (\zeta )= \left [\zeta - h_\infty \right ]^{-1}$. \
Furthermore, making use of the same arguments as in \cite[\S 3.2]{Al},
it follows that for any rapidly decreasing test function $\varphi $    

\be
\lim_{\nu \to \infty } \left \langle \varphi,\left [ r_\nu (\zeta ) -
    r_\infty (\zeta ) \right ] \varphi \right \rangle_{L^2 (\Real)} 
=0, \ \Im \zeta \ge 0 , \label {doublelimit} 
\ee
uniformly for $\zeta $ belonging to a compact set $[a,b]\times i[0,c]$
for any $a,b,c >0$.  

\subsubsection* {Computation of the coefficient $a$}

For $\nu\to \infty$ we have
\be
\sigma_d (h_\nu) = \left \{ \mu_1 = -\frac 14 + O(\nu^{-1}) \right \} 
\, .
\ee
We take $\nu$ large enough so that for the set $\Sigma = \left \{ E
  =E_{j,n} = \mu_j + E_n \right \}$ of eigenvalues of $H_0^{V_\nu}$
holds  
\be
E_{1,1} = E_1 - \frac 14 + O (\nu^{-1}) < E_1 < E_{1,2} = E_2 - \frac
14 + O (\nu^{-1}) < E_2 . 
\ee
Then we apply the perturbative theory to the embedded eigenvalue
\be
E=E_{1,2} =E_2 + \mu_1
\ee
with the associated eigenvector
\be
\psi (x,y,z)= \varphi_1 (x) \chi_2 (y,z)\, .
\ee
In such a case $k^{\star} =2$ and 
\be 
\label{imag}
&& \Im a_2 = \\
&& -\lim_{\rho \to 0^+} \Im \left \{\sum_{k=1,2} \langle d_k, r_\nu
  (E-E_k - i \rho ) d_k \rangle_{L^2 (\Real )} - \left | \left \langle
      \psi , U_{\eps}^{V_\nu} \psi \right \rangle_{L^2 (\Omega }
  \right |^2 (i\rho )^{-1} \right \} 
\nonumber
\ee
where $r_\nu (\zeta )$ is the resolvent of $h_\nu $ and 
\be \label{d_k} 
 d_k (x) = \langle \chi_k,
U_{\eps}^{V_\nu} \psi \rangle_{L^2 (\omega )} 
\ee

An integration by parts shows that $ \langle \chi_2 , \pd_\tau \chi_2
\rangle_{L^2 (\omega )} =0$. \ We thus get   
\bee
\langle \psi , U_\eps^{V_\nu} \psi \rangle_{L^2 (\Omega )} &=& -2\eps
\langle \varphi_1 , \dot \alpha \pd_x \varphi_1 \rangle_{L^2 (\Real )} \langle
\chi_2 , \pd_\tau \chi_2 \rangle_{L^2 (\omega )}  \\ 
&& -\eps \langle \varphi_1 , \ddot \alpha \varphi_1 \rangle_{L^2 (\Real )}
\langle \chi_2 , \pd_\tau \chi_2 \rangle_{L^2 (\omega )} 
\\ && -\eps^2 \langle \varphi_1 , \dot \alpha^2 \varphi_1 \rangle_{L^2 (\Real )}
\langle \chi_2 , \pd_\tau^2 \chi_2 \rangle_{L^2 (\omega )} \\ 
&=& - C_0 \eps^2\, ,
\eee
where
\bee
C_0= \langle \varphi_1 , \dot \alpha^2 \varphi_1 \rangle_{L^2 (\Real )}
 \langle \chi_2 , \pd_\tau^2 \chi_2 \rangle_{L^2 (\omega )} = \langle \chi_2 , \pd_\tau^2 \chi_2 \rangle_{L^2 (\omega )} \, .
\eee

Furthermore, from (\ref{d_k}) and we obtain that 
\be
\label {dkbis}
d_1 (x)= -2\eps C_1\, \pd_x \varphi_1 - \eps^2 C_2\, \varphi_1, \quad
d_2(x)= -\eps^2 C_0\, \varphi_1   
\ee
where
\bee
C_1 = \langle \chi_1, \pd_\tau \chi_2 \rangle_{L^2 (\omega )}\, \ \
\mbox { and } \ \ C_2 = \langle \chi_1, \pd_\tau^2 \chi_2 \rangle_{L^2
  (\omega )}\, . 
\eee
Collecting all these facts and keeping in mind that $E=E_2+\mu_1$ and
$\varphi_1$ is the eigenfunction of $h_\nu$ with eigenvalue $\mu_1$ we
get  
after some tedious, but straightforward calculations, that
\bee
\lim_{\rho\to 0+}\Im\left \{\langle d_2 , r_\nu (E-E_2-i \rho ) d_2
  \rangle_{L^2(\Real)} - \left |\langle \psi , U_{\eps}^{V_\nu} \psi
    \rangle \right|_{L^2(\Omega)}^2(i\rho)^{-1} \right\}= 0 \, . 
\eee
This implies
\be \label{im-a2} 
\Im a_2 & = & -\lim_{\rho \to 0^+} \Im \langle d_1 , r_\nu( E-E_1-i 
\rho ) d_1 \rangle_{L^2(\Real)}  \\ 
& = & -\lim_{\rho \to 0^+} \Im\left[4 \eps^2 |C_1|^2 \langle \pd_x 
\varphi_1 , r_\nu ( E - E_1 -i \rho ) \pd_x \varphi_1 \rangle\right ] + 
O(\eps^3 )\, .
\ee
We now pass to the limit $\nu\to\infty$ which implies 
\be \label{limit} 
\mu_1 \to - \frac {1}{4}\, ,\quad \varphi_1 \to \phi = \sqrt {\frac 
{1}{2}}\, \, e^{-|x|/2} \, , \quad  h_\nu \to h_\infty \, ,
\ee
where the last limit is reached in the norm resolvent sense. \ We
recall also that the resolvent $[\zeta-h_\infty]^{-1}$ has the kernel
given by 
\bee
\K_\zeta (x,x') = \K^0_\zeta (x,x') + \K^1_\zeta (x,x')\, ,
\eee
where
\bee
\K^0_\zeta (x,x') = \frac {1}{2ki}\, e^{ik |x-x'|} \, ,\quad
\K^1_\zeta (x,x') = -\frac {1}{2k} \frac {1}{2k+i}\,  e^{ik
  [|x|+|x'|]}, \ \ \zeta =k^2, \, \Im k >0\, , 
\eee
see \cite{Al}. In our case
\be
\zeta= k^2 =  E-E_1-i\rho = E_2-E_1+\mu_1 -i\rho\, .
\ee
We will denote by $\K_\zeta^0$ and $\K_\zeta^1$ the integral operators
with the kernels $\K^0_\zeta (x,x')$ and $\K^1_\zeta (x,x')$
respectively. \ Note that $\pd_x \varphi_1(x)$ is an odd function,
which implies that $\K^1_\zeta \pd_x \varphi_1 \equiv 0$. \ Now $h_\nu
\to h_\infty $ as in (\ref {doublelimit}). \ Since $E-E_1$ is not an
eigenvalue of $h_\nu $ for any $\nu $ large enough (in fact $E-E_1$
belongs to the absolute continuous spectrum of the operators $h_\nu$
and $h_\infty$) and $d_1$ is an exponentially decreasing function as
$|x|\to \infty$, we can pass to the limit $\nu\to\infty$ replacing
$r_\nu ( E - E_1 -i \rho)$ on the r.h.s.~of (\ref{im-a2}) by
$\K_\zeta^0$ and $\varphi_1$ by $\phi$:   
\be
\label {limitedoppio}
\lim_{\nu\to\infty}\, \Im a_2 &=&  - \lim_{\nu \to \infty } \lim_{\rho
  \to 0^+} \Im \langle d_1 , r_\nu( E-E_1 - i \rho )\, d_1 
\rangle_{L^2(\Real)}  \\  
& = & -\lim_{\rho \to 0^+} \Im\left[4 \eps^2 |C_1|^2 \langle \pd_x
  \phi , r_\infty ( E - E_1 -i \rho )\, \pd_x \phi \rangle\right ] +
O(\eps^3 ), \nonumber 
\ee
where the remainder term is uniform with respect to $\rho$. \ An
explicit computation then gives  
\be \label{rho}
\lim_{\rho \to 0^+} \Im \langle \pd_x \phi , \K_\zeta^0\, \pd_x \phi
\rangle = \frac {4 \sqrt {E_2-E_1 +\mu_1 }}{\left [1+4(E_2-E_1 +\mu_1)
  \right ]^2} \, . 
\ee 
In view of  \eqref{limit} and \eqref{rho} we get
\bee
\lim_{\nu\to\infty}\, \Im a_2 = -\eps^2 |C_1|^2\, \frac {\sqrt
  {E_2-E_1 -\frac 14 }}{(E_2-E_1)^2 } \, .   
\eee
The proof is complete.

\end{document}